\documentclass[journal]{IEEEtran}
\usepackage{cite}
\usepackage{amsmath,amssymb,amsfonts}
\usepackage{algpseudocode}
\usepackage{algorithm}
\usepackage{graphicx}
\graphicspath {{./img}}
\usepackage{textcomp}
\usepackage{cuted}
\usepackage{capt-of}

\usepackage[hyphens]{url}
\usepackage[hidelinks]{hyperref}
\newif\ifinpaper
\inpapertrue
\newcommand{\inpaper}[2]{\ifinpaper #1\else #2\fi}

\def\BibTeX{{\rm B\kern-.05em{\sc i\kern-.025em b}\kern-.08em
    T\kern-.1667em\lower.7ex\hbox{E}\kern-.125emX}}
\begin{document}
\title{Sugar Shack 4.0: Practical Demonstration of an IIoT-Based Event-Driven Automation System}
\author{Thomas Bernard, \IEEEmembership{Student member, IEEE}, François Grondin, \IEEEmembership{Member, IEEE}, and Jean-Michel Lavoie
\thanks{This work was supported in part by the Natural Sciences and Engineering Research Council of Canada (NSERC) and the Consortium de recherche et innovations en bioprocédés industriels au Québec (CRIBIQ). \textit{Corresponding author: Thomas Bernard.}}
\thanks{Thomas Bernard and François Grondin are with the Department of Electrical and Computer Engineering, Université de Sherbrooke, Sherbrooke, QC J1K 2R1, Canada. (e-mail: thomas.bernard@usherbrooke.ca, francois.grondin2@usherbrooke.ca).}
\thanks{Jean-Michel Lavoie is with the Department of Chemical and Biotechnological Engineering, Université de Sherbrooke, Sherbrooke, QC J1K 2R1, Canada. (e-mail: jean-michel.lavoie2@usherbrooke.ca).}}

\markboth{}{}

\maketitle

\begin{abstract}
    This paper presents a practical alternative to programmable-logic-controller-centric automation by implementing an event-driven architecture built with industrial Internet of Things tools. A layered design on a local edge server (i) abstracts actuators, (ii) enforces mutual exclusion of shared physical resources through an interlock with priority queueing, (iii) composes deterministic singular operations, and (iv) orchestrates complete workflows as state machines in Node-RED, with communication over MQTT. The device layer uses low-cost ESP32-based gateways to interface sensors and actuators, while all automation logic is offloaded to the server side. 
As part of a larger project involving the first scientifically-documented integration of Industry 4.0 technologies in a maple syrup boiling center, this work demonstrates the deployment of the proposed system as a case-study. Evaluation over an entire production season shows median message time of flight around one tenth of a second, command issuance-to-motion latencies of about two to three seconds, and command completion near six seconds dominated by actuator mechanics; operation runtimes span tens of seconds to minutes. These results indicate that network and orchestration overheads are negligible relative to process dynamics, enabling modular, distributed control without compromising determinism or fault isolation. The approach reduces material and integration effort, supports portable containerized deployment, and naturally enables an edge/cloud split in which persistence and analytics are offloaded while automation remains at the edge.
\end{abstract}


\section{Introduction}
\label{sec:introduction}
\begin{figure}
        \centering
        \includegraphics[width=\inpaper{\linewidth}{0.95\linewidth}]{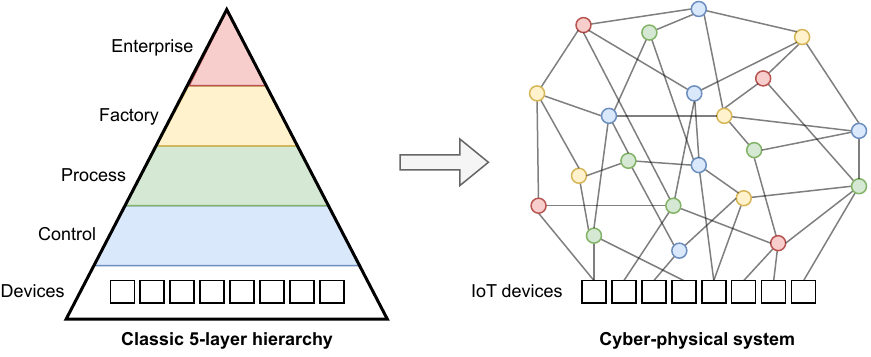}
        \caption{Architectural transition of industrial automation systems}
        \label{fig:cps}
    \end{figure}

    \inpaper{
        \IEEEPARstart{P}{rogrammable}
    }{
        Programmable
    }
    Logic Controllers (PLCs) remain the foundation of industrial control, yet the decentralization and connectivity promoted by Industry~4.0 strain the fit of traditional automation systems and long-standing IEC~61131 programming models. 
    Industry is experiencing a shift away from the traditional 5-layer \textit{automation pyramid} towards ubiquitously connected Cyber-Physical Systems (CPS), as shown in Fig.~\ref{fig:cps}. Legacy architecture is generally criticized for its unitaleral upwards data flow, where each jump between layers add information compression and latency, as well as limit the capability of upper layers to provide timely contributions at the automation level~\cite{frereIndustry40Germany2018}. In this context, recent surveys highlight integration and agility challenges for adapting PLC-centric stacks to the requirements of Industry 4.0~\cite{sehrProgrammableLogicControllers2021}. Large enterprises have pursued this transition via substantial retrofits to integrate legacy systems with cloud and analytics platforms, but the capital intensity and skill burden remain significant barriers, particularly for small and medium-sized businesses (SMBs). Analyses of cloud adoption in manufacturing emphasize that OPEX-oriented, pay-as-you-go models offer a more accessible solution by lowering adoption thresholds~\cite{reznikovEconomicImpactCloud2023}. 

    Internet of Things (IoT) technologies are now commonplace in consumer applications, such as smart homes, for both data acquisition and actuator control \cite{arzoTheoreticalDiscussionSurvey2021}. In contrast, Industrial IoT (IIoT) applications operate under tighter real-time, interoperability, security, and hardware constraints. This is a factor explaining why IIoT has been widely adopted for monitoring and optimization (where it is present \textit{alongside} automation systems) while its use as a direct vehicle for distributed automation remains comparatively limited~\cite{paniaguaIndustrialFrameworksInternet2021}.

    An inherent challenge with networked/distributed control systems in an industrial context is determinism, since they introduce additional sources of variability compared to traditional fixed-rate PLC systems. Indeed, wireless devices are attractive for flexibility and scalability but demand careful attention to latency/jitter and reliability in high-performance settings~\cite{pangWirelessHighPerformanceCommunications2017}. In this regard, methodologies for reliable IIoT-based distributed automation argue that functional equivalence to centralized control can be preserved under realistic communication assumptions, provided message loss and timing variability are explicitly accounted for and appropriate fault-handling mechanisms are implemented~\cite{lesiReliableIndustrialIoTbased2019}.
    
    Many industrial process operations are long-running and prioritize consistent execution and safety over millisecond actuation deadlines (e.g., pasteurization profiles on seconds-minutes horizons)~\cite{burkeDairyIndustryProcess2018}, which presents an opportunity to evaluate trading off strict timing constraints in favor of modularity and scalability, while conforming to process specifications.

    This paper contributes a practical, IIoT-based approach to distributed automation that implements field devices as sensor and actuator gateways and offloads all automation logic to a local server. It introduces two main contributions: (i) a modular device abstraction layer that decouples control logic from hardware implementation and enables horizontal scaling and rapid reconfiguration; and (ii) an asynchronous interlock with priority-aware queueing that shifts concurrency handling from design-time enumeration to safe, reusable and deterministic runtime arbitration. This work is part of a larger project aiming to optimize the productivity and efficiency of a large-scale maple syrup production facility using Industry 4.0 technologies, of which elements are presented as a case study supporting the viability of the proposed approach in a traditional industry with modest resources.

    The remainder of the paper is structured as follows: Section \ref{sec:related-work} reviews the related work leading to IIoT-based automation, Section \ref{sec:system-architecture} details the server-side software architecture, Section \ref{sec:implementation-details} presents the implementation as part of the case study and Section \ref{sec:evaluation-results} evaluates the performance of the proposed system.

\section{Related work}
\label{sec:related-work}

    \subsection{PLC transformation in the 4.0 era} 
        A comprehensive review positions PLCs as robust yet structurally misaligned with Industry 4.0, weighing limitations of the IEC 61131 model around event-driven control and scalable distributed deployment. Poor encapsulation/abstraction leads to costly prototype-and-test iterations, and untimestamped inter-component messaging hinders synchronization~\cite{sehrProgrammableLogicControllers2021}.
        Advances in PLC software engineering (e.g., object-oriented and model-driven approaches) aim to address this issue by raising abstraction and reuse; for instance, object-oriented IEC~61131 design has been reported to improve understandability and maintainability in user studies compared to conventional implementations~\cite{obermeierModelDrivenApproachObjectOriented2015}. Semantic models (e.g., OntoPLC) target code exchange and reuse across vendors/projects~\cite{anOntoPLCSemanticModel2021}.

        Early service-oriented architectures (SOA) for factories integrated legacy field devices and controllers via adapters and service gateways, exposing device capabilities to higher-level orchestration while preserving installed assets~\cite{karnouskosIntegrationLegacyDevices2009}. This established that service interfaces could wrap existing control points; however, the primary focus was interoperability and vertical integration rather than network-native, distributed actuation with resource coordination across peers.

        Virtualized PLCs hosted in the cloud demonstrated feasibility and cost flexibility but documented performance and predictability caveats versus dedicated hardware; comparative evaluations highlighted accessibility/security benefits alongside minor ($\approx$ 3 ms) timing trade-offs~\cite{givehchiControlasaserviceCloudCase2014}. 
        Complementary \emph{industrial automation as a cloud service} work introduced delay-tolerant control loops with compensators for network variability (soft real time), showcasing the viability of full cloud placement to integrate three sites across different continents~\cite{hegazyIndustrialAutomationCloud2015}. 
        The \emph{PLC as a smart service} concept further presents middleware for distributing controller instances, reporting a reliable implementation for controlling processes where reaction times up to 150 ms are acceptable~\cite{langmannPLCSmartService2019}. 

        Quantified evaluations of \textit{control-as-a-service} under microservice orchestration report closed-loop feasibility with measured time budgets: in a DC-motor loop, the DAQ→PID→actuate round-trip averaged 189.6 ms (jitter 32 ms; worst 354.5 ms), within a 300 ms cycle~\cite{bighetiControlServiceMicroservice2019}.
        In a pilot pipe-pressure loop, an API-gateway-orchestrated pipeline was instrumented via a tracer service, collecting 133 samples per service (399 for the gateway); mean end-to-end orchestration time was 62 ms over Wi-Fi and 51 ms on Ethernet~\cite{pontarolliMicroserviceOrchestrationProcess2020}. 
        Complementing orchestration with reconfigurability, a containerized IoT-PLC demonstrated runtime controller switching and live migration with checkpoint times of 9.7 s (PID), 7.7 s (MPC), 17.7 s (manager), and 53.9 s (HMI), while loop and control latencies remained statistically comparable to a non-container baseline~\cite{melladoDesignIoTPLCContainerized2022}.

        Overall, evidence shows that networked automation is viable in meeting timing budgets. However, most challenges brought forward in~\cite{sehrProgrammableLogicControllers2021} are still valid concerns in approaches where PLCs are virtualized and/or distributed. This leaves an opportunity to investigate beyond the PLC; into event-driven, IIoT-based systems with explicit timing, especially when employing distributed devices around shared physical assets.

    \subsection{IIoT-based platforms}
        IIoT platforms such as Arrowhead-based systems demonstrate dynamic service compositions, device provisioning, and monitoring pipelines with reconfiguration capabilities that scale to thousands of concurrent data flows, while safeguarding availability and latency in production settings~\cite{hastbackaDynamicEdgeCloud2022}. Complementing this, local “automation clouds” show that colocated service hubs can encapsulate legacy field assets and enforce interoperability and security; case studies report reductions in manual integration effort between 3 and 5 times when compared to bespoke point-to-point integrations~\cite{delsingEnablingIoTAutomation2016}. At a higher level, surveys converge on softwarization, microservices, containerization, and cloudification as the enabling stack for automated orchestration at scale~\cite{arzoTheoreticalDiscussionSurvey2021}.

        On the messaging and orchestration plane, MQTT has emerged as the default lightweight backbone: a brokered publish/subscribe overlay with QoS delivery guarantees, retained messages, and ``last will'' signals that support state convergence and failure detection. Practical automation stacks use MQTT to translate traditional protocol data units and bridge legacy subsystems into IIoT topics, yielding scalable, interoperable data paths for monitoring and decision support~\cite{yolovDevelopmentIntegrationIndustrial2025}. At the application level, Node-RED provides event-driven flow composition that is already sufficient for supervisory functions: open implementations combine Modbus~TCP to interface local PLCs, MQTT for higher-layer exchange, and time-series persistence (e.g., InfluxDB), with experiments reporting near-instantaneous end-to-end updates and low bandwidth footprints under QoS-tested scenarios~\cite{nitulescuSupervisoryControlData2020}. Device-cloud gateways further validate feasibility at the edge: microcontroller implementations simultaneously handle analog 4-20\,mA I/O and Modbus-RTU while streaming four sensor channels and actuating four outputs in real time to the cloud via WebSocket, including tests with six Modbus power meters~\cite{nuratchIIoTDevicesCloud2017}. Beyond supervision, microservices-based cloud-edge condition monitoring reaches production-relevant outcomes (e.g., $\approx$90\% diagnostic accuracy with a $\approx$50\% prediction-error reduction), underscoring the robustness of the data/management plane~\cite{yangMicroservicesbasedCloudedgeCollaborative2022}. 

        The literature validates scalable discovery, messaging, and lifecycle management for IIoT-based automation; however, most platforms are focused on data acquisition and management and stop short of natively integrating distributed actuators around shared physical resources with explicit arbitration and event-driven control. The present work addresses this gap by implementing such a control plane with industry-proven IIoT tools: MQTT and Node-RED.

\section{Software system architecture}
\label{sec:system-architecture}

    The proposed automation system is organized into a layered architecture that separates device control from higher-level sequencing and coordination.
    At the foundation, an \textbf{actuator abstraction} integrates the control and feedback behavior of each type of actuator employed in the process.  
    These abstractions are used by a \textbf{group command} wrapper that serializes any combination of actuator commands into a single logical action.
    Alongside commands, a \textbf{resource interlock and operation queueing} mechanism ensures safe, exclusive access to shared physical resources in the industrial process. 
    \textbf{Singular operations} then encapsulate those two concepts alongside sensor condition fulfillment into reusable sequences, which are then called from higher-level \textbf{automation routines} representing complete process workflows. 

    Control logic is implemented in Node-RED to enforce a strictly event-driven runtime. Encapsulation relies on "link call" nodes to implement the layered architecture: a caller passes a payload to a subflow with a generated \verb|_msgId|, and flow state is restored upon return with the same \verb|_msgId| by the callee. Each call node carries a configurable timeout that enables fault handling. 
    Shared state that must be persisted across concurrent calls is kept in context storage (flow/global). With the focus of keeping flows self-explanatory, the remainder of this section favors showing flow snippets taken directly from the implementation as examples.

    \subsection{Actuator abstractions}
    \label{sec:actuator-abstractions}

\begin{figure}
            \centering
            \includegraphics[width=\inpaper{\linewidth}{0.95\linewidth}]{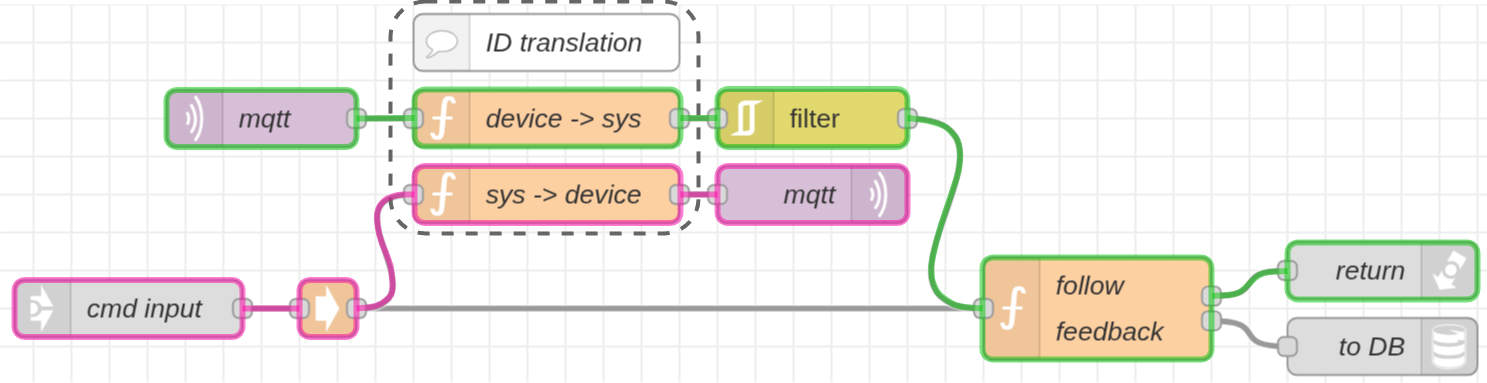}
            \caption{\textbf{Valve actuator wrapper.} Device- vs. system-level ID mapping is circled in dotted line, "command input to publish path" is highlighted in pink, and "feedback to return path" is highlighted in green.}
            \label{fig:valve_cmd}
        \end{figure}

        At the hardware boundary, an \emph{abstraction layer} encapsulates the behavior of each actuator type into a uniform interface.
        In the present case-study, this includes valves and pumps. 
        This layer flows information from two paths: (i) a command input called from the layer above, that leads to an MQTT publish to the device, and (ii) a feedback input that is received from the device via an MQTT subscription and leads to a return to the caller.
        Each actuator has a unique ID at the system level, which is mapped to and from a corresponding IO number at the device level (used in MQTT topics). 
        Fig.~\ref{fig:valve_cmd} shows an example of the valve actuator wrapper used in the case-study.
        When a command is dispatched, its state is updated in a global context object, marking (i) its target value (ii) its state as "pending", and (iii) the command input's \verb|_msgId|. Feedback is received periodically for all actuators and serialized into an individual message for each system-level ID, applying analog tolerance thresholds to ensure adequate operation. A \textit{report-by-exception} filter passes only messages with a changed value per each ID, avoiding repeated messages for inactive commands. Completing the example of a valve command: its state is marked as "moving" on the first message indicating departure from the initial position; and marked as "stopped" when the target position is reached, returning the command call.

\begin{figure}[H]
            \centering
            \includegraphics[width=\inpaper{\linewidth}{0.95\linewidth}]{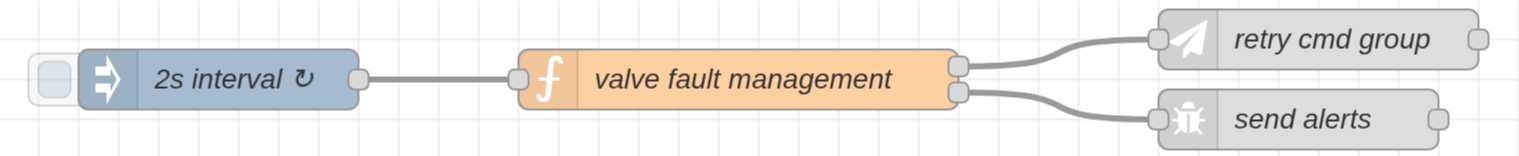}
            \caption{Valve fault management.}
            \label{fig:valve_fault}
        \end{figure}

        On this layer, a fault handler periodically monitors the behavior of all actuators of one type against configurable failure modes and thresholds. Fault type and relevant timing data are stored in the same global object as the actuator state. The handler can dispatch a retry command group (\S\ref{sec:actuator-command-grouping}) as needed to account for network unreliability, as well as send error messages to communication channels. An escalated error implies that a given command won't return to its caller; the fault is then handled by the timeout mechanism of the layer above. Fig.~\ref{fig:valve_fault} shows an example of the valve fault handler.

    \subsection{Actuator command grouping}
    \label{sec:actuator-command-grouping}

\begin{figure}[H]
        \centering
        \includegraphics[width=\inpaper{\linewidth}{0.95\linewidth}]{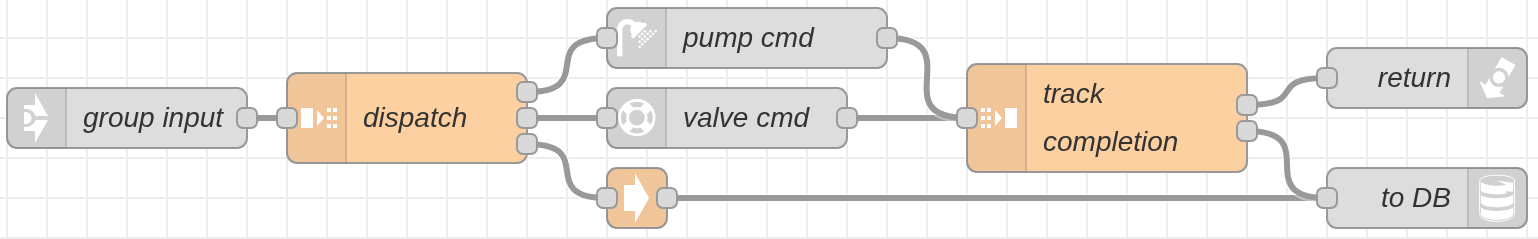}
        \caption{Group command flow.}
        \label{fig:group_cmd}
        \end{figure}

        Actuator command grouping implements a single point of entry for actuator control, wrapping every actuator abstraction. It expects a JSON payload with an array of key-value pairs for each actuator type, which is serialized into individual command calls to the actuator abstraction layer. 
        Each active group command is stored in a global context object with its \verb|_msgId|, along with the pending actuator commands to track. 
        The group's \verb|_msgId| is also passed down into the message to the actuator abstraction layer, to be passed back up upon command return for completion tracking.
        When all individual commands have returned, the group command returns to its caller with a success message. No timeouts nor fault handling is implemented at this layer; timeouts are defined in the layer above (\S\ref{sec:singular-operations}) on a per-operation basis.
        Fig.~\ref{fig:group_cmd} shows an example of the group command flow in the present case-study.

        \inpaper{}{\clearpage}

    \subsection{Resource interlock and operation queueing}
    \label{sec:interlock}

        The system implements an asynchronous interlock layer that ensures \emph{mutual exclusion} over shared process resources. Operations (\S\ref{sec:singular-operations}) request atomic\footnote{Node-RED's underlying Node.js event loop implementation is single-threaded and non-preemptive; it allows asynchronous execution for branching flows but guarantees atomic execution for any given \textit{function node}~\cite{nickolearyMakingFlowsAsynchronous2019}.} acquisition of a set of one or more resource(s): if the resources are free, the request is granted immediately; otherwise requests are enqueued with priority scheduling. Releases and faults trigger re-evaluation of the queue. The objective of this subsystem is to allow the development of functional operations that use overlapping physical resources without having to explicitly plan every possible conflicting interaction at design time. In the present case-study, the shared resources are sections of piping separated schematically into five colored "lines"; several use-cases employ more than one line at the same time. The following algorithm set defines the behavior of the interlock system.

        \textbf{Invariants}:
        \begin{itemize}
            \item $R$ -- shared resource set.
            \item $P[op\_id]$ -- operation priorities, unique by $op\_id$.
        \end{itemize}
        
        \textbf{State variables} (in Node-RED global context):
        \begin{itemize}
            \item $O[r \in R] \in \{\texttt{null}, \texttt{fault}, op\_id\}$ -- \\resource ownership (\texttt{fault} blocks all operations).
            \item $Q = \{(op\_id, R' \subseteq R, p = P[op\_id], \verb|_msgId|)\}$ -- \\operation queue, unique by $op\_id$.
        \end{itemize}

        \begin{algorithm}[H]
            \caption{Acquire$(op\_id,\, R' \subseteq R)$}
            \begin{algorithmic}
                \If{$\forall r \in R': O[r] = null$}
                    \ForAll{$r \in R'$}
                        \State $O[r] \gets op\_id$
                    \EndFor
                    \State \Return \verb|_msgId|
                \ElsIf{$op\_id \notin Q$}
                    \State enqueue$(op\_id, R', P[op\_id], \verb|_msgId|)$
                \EndIf
            \end{algorithmic}
        \end{algorithm}

        \begin{algorithm}[H]
            \caption{Release($op\_id$)}
            \begin{algorithmic}
                \State $released \gets \emptyset;\quad granted[] \gets \emptyset$
                \ForAll{$r \in R$}
                \If{$O[r] = op\_id$}
                    \State $O[r] \gets null$ 
                \EndIf
                \EndFor
                \State $released \gets \verb|_msgId|$
                \State sort $Q$ by $p$ descending
                \ForAll{q $\in Q$}
                \If{$\forall r \in q.R': O[r] = null$}
                    \ForAll{$r \in q.R'$}
                        \State $O[r] \gets q.op\_id$
                    \EndFor
                    \State remove $q$ from $Q$
                    \State add $q.\verb|_msgId|$ to $granted$
                \EndIf
                \EndFor
                \State \Return $released$ and $granted$ (serialized)
            \end{algorithmic}
        \end{algorithm}

        Adding to the normal behavior to \textit{acquire} and \textit{release} resources, the interlock system also implements a fault isolating mechanism. It is called in the same way as a \textit{release} request, with the purpose of locking the resources involved in the fault. 
        Each resource has a predefined \textit{lockout group command} that is issued to all relevant actuators when the fault is triggered. For example, in the case-study, this involves closing all valves and stopping the pump in the affected section.
        When a fault call is issued: $O[r] \gets fault$ for all $r \in R'$ owned by $op\_id$, the lockout group command is dispatched for all $r \in R'$, and a message is sent to the appropriate communication channels for troubleshooting.
        This prevents operations from re-using the faulted resources until the cause is resolved by a manual intervention, while not impeding system operation in unaffected parts of the process.
        The system also periodically checks the elapsed time for queued requests and active operations against parameterized thresholds, sending preventive warning messages for unusual behavior.

\begin{figure}[H]
        \centering
        \includegraphics[width=\inpaper{\linewidth}{0.95\linewidth}]{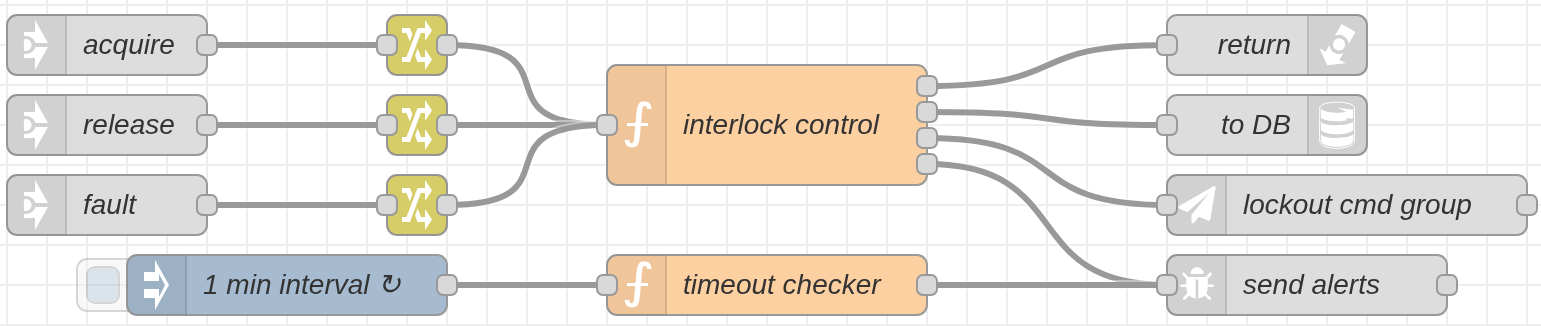}
        \caption{Interlock control flow}
        \label{fig:interlock}
        \end{figure}

        As for \S\ref{sec:actuator-abstractions} and \S\ref{sec:actuator-command-grouping}, the interlock system leverages Node-RED link call nodes to preserve flow execution. Since \textit{Acquire}, \textit{Release} and \textit{Fault} requests are called from different points, the \verb|_msgId| from each call is used to return in the right place. This happens either (i) immediately from a granted \textit{Acquire} call or any \textit{Release}/\textit{Fault} call (as part of a single \textit{function node} execution; the right \verb|_msgId| is already attached to the message), or (ii) by attaching the \verb|_msgId| saved in context from a queued \textit{Acquire} call. In the case of a \textit{Release} call that triggers the return of one or more queued \textit{Acquire} call(s), several return messages are serialized, each one with the unique \verb|_msgId| from their respective source. Fig.~\ref{fig:interlock} shows the implementation of the interlock system.

    \subsection{Singular operations}
    \label{sec:singular-operations}

        Building upon the actuator abstraction and interlock mechanisms, the system defines \emph{singular operations} as modular building blocks that encapsulate reusable sequences, presenting the following structure:

        \begin{enumerate}
            \item \textbf{Lock acquisition}: the operation requests exclusive access to the required resources through the interlock.
            \item \label{it:group-commands}\textbf{Group command(s)}: once access is granted, the operation issues one or more group commands in sequence via the actuator abstraction layer.
            \item \label{it:condition-fulfillment}\textbf{Condition fulfillment}: the operation monitors process state from relevant sensors\footnotemark[1] until the fulfillment condition is met. Steps \ref{it:group-commands} and \ref{it:condition-fulfillment} can be repeated in any combination depending on the application.
            \item \textbf{Lock release}: before returning, all resources are released from the interlock system.
        \end{enumerate}
        \footnotetext[1]{Deployed sensors are presented in \S\ref{subsec:instrumentation}. Live operation-relevant values are forked from MQTT subscriptions into global context storage.}

        \begin{figure*}[t]
            \centering
            \includegraphics[width=\linewidth]{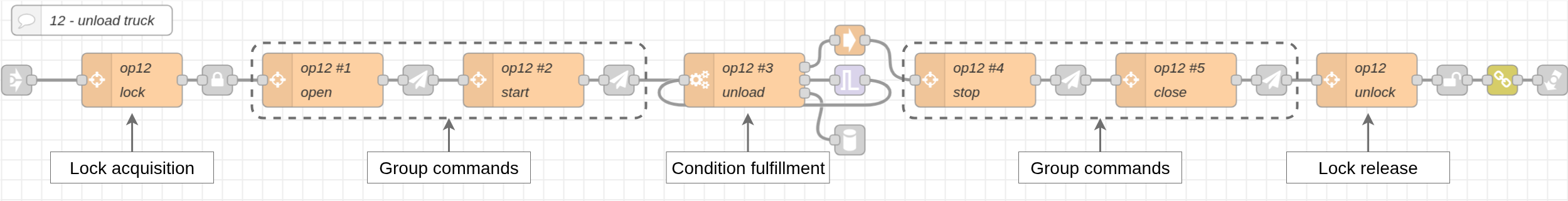}
            \captionof{figure}{Annotated singular operation example (op12)}
            \label{fig:singular-ops-example}
        \end{figure*}

        Fig.~\ref{fig:singular-ops-example} shows an example implementation of a singular operation from the case-study.

        By definition, singular operations are deterministic and are meant to be fully executed in a single call, implying that they should be as short as possible while fulfilling the following conditions:
        \begin{enumerate}
            \item[\textbf{C1.}] only a single instance of a given operation can be executed at a time;
            \item[\textbf{C2.}] an operation can only command actuators that are strictly included within the boundary of acquired resources;
            \item[\textbf{C3.}] prior to releasing resources, the operation must return all involved actuators to their original (idle) state.
        \end{enumerate}

\begin{figure}[h!]
			\centering
			\includegraphics[width=\inpaper{\linewidth}{0.95\linewidth}]{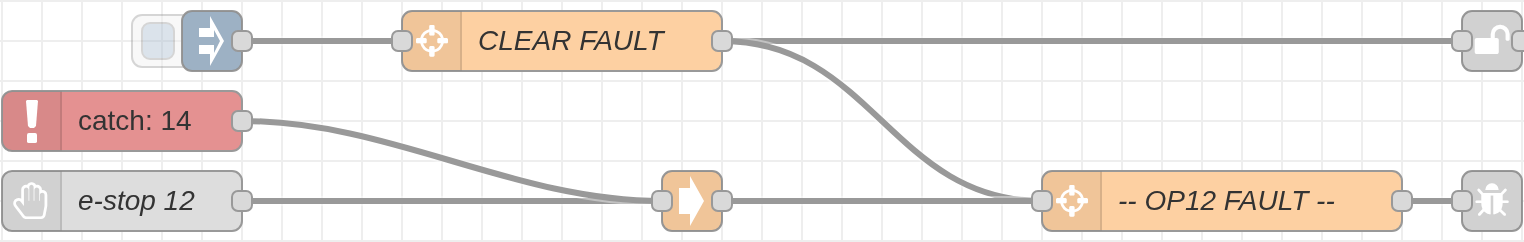}
			\caption{Example of fault handling for op12}
			\label{fig:op12_fault}
		\end{figure}

        The most important part of the fault handling subsystem is implemented at the singular operation level. Every link call node (small grey nodes in Fig.~\ref{fig:singular-ops-example}) has a predefined timeout value that throws a Node-RED error when exceeded. These errors, combined with those that can be thrown from within a "Condition fulfillment" function node, are caught by a \textit{catch} node that leads into faulting the operation at the interlock layer, as examplified in Fig.~\ref{fig:op12_fault}. Where applicable, a fault can also be manually triggered by an emergency stop HMI button (which is out of scope for this paper). When the fault state is entered, the operation and relevant actuators are locked out until the fault is cleared by a manual intervention.

\begin{figure}[h!]
			\centering
			\includegraphics[width=\inpaper{0.93\linewidth}{0.90\linewidth}]{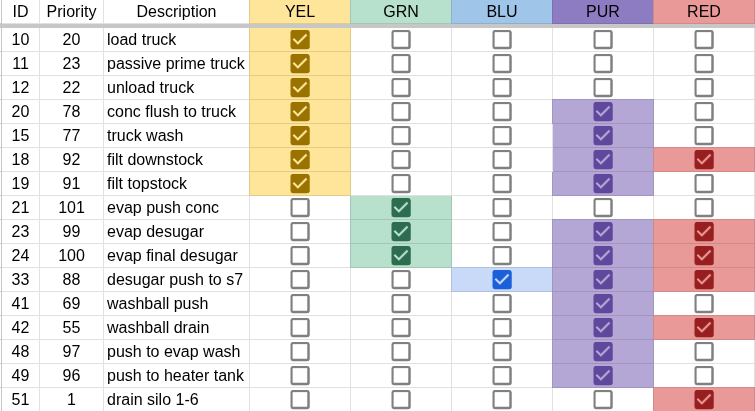}
			\caption{List of singular operations used in the case-study}
			\label{fig:op_table}
		\end{figure}

        To complement the examples given and the shared resource application mentioned in \S\ref{sec:interlock}, Fig.~\ref{fig:op_table} shows the list of singular operations used in the case-study.

        \inpaper{}{\clearpage}

    \subsection{Automation Routines}

        At the highest level of the architecture, \emph{automation routines} coordinate singular operations into process workflows. 
        Each routine is modeled as a state machine that evaluates conditions periodically from context storage or reacts to external triggers. 
        During operation, the routine dispatches at most one operation from its set of predefined steps. 
        The return of the operation call updates the routine's state, thereby enabling subsequent transitions. 
        This design ensures that only a single operation can be pending within a routine at any time, avoiding concurrency conflicts at the routine level.

        Formally, a routine can be expressed as a tuple
        \[
        R = (S, s_0, O, T)
        \]
        where $S$ is the set of states, $s_0 \in S$ is the initial state, $O$ is the set of operations assigned uniquely to $R$, and $T \subseteq S \times C \times O \times S$ defines transitions conditioned on trigger $C$ that dispatch operation $o \in O$ and move to a new state. 
        The uniqueness of $O$ across routines enforces strict ownership of operations (i.e., an operation identifier cannot be invoked by multiple routines, which ensures fulfillment of condition \textbf{C1} mentioned in \S\ref{sec:singular-operations}).

        \begin{algorithm}[H]
        \caption{Routine Execution}
        \begin{algorithmic}
        \State $s \gets s_0$
        \While{routine active}
        \If{trigger condition $C$ satisfied in state $s$}
            \State dispatch corresponding operation $o$
            \State wait until $o$ returns completion
            \State $s \gets$ next state according to $T$
        \EndIf
        \EndWhile
        \end{algorithmic}
        \end{algorithm}

\begin{figure}[h!]
			\centering
			\includegraphics[width=\inpaper{0.99\linewidth}{0.95\linewidth}]{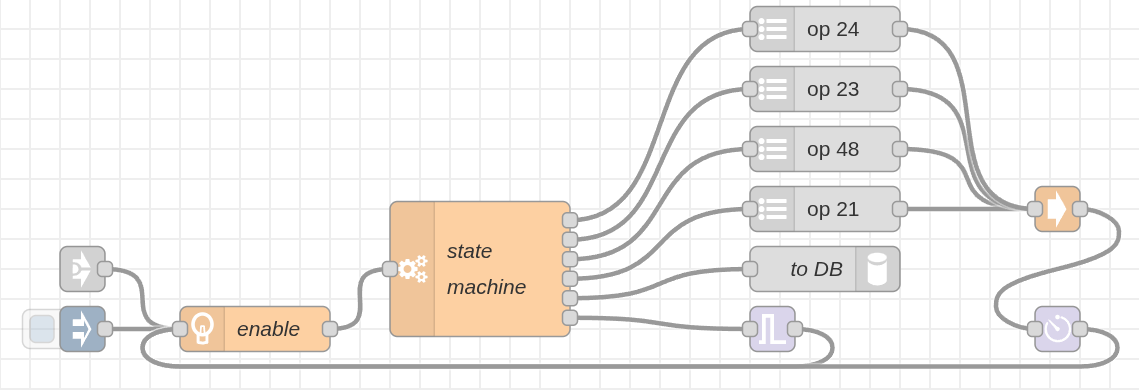}
			\caption{Example of a routine}
			\label{fig:routine}
		\end{figure}

        Fig.~\ref{fig:routine} shows an example of a routine. Tineouts and error-handling at this level are implemented ad-hoc; since operations have per-call timeouts, a global timeout is not relevant. Nevertheless, the state machine node has the possibility of throwing errors when relevant for the application.

\section{Case-study implementation details}
\label{sec:implementation-details}

    The proposed architecture was deployed and validated in a large-scale maple syrup production facility. 
    This section details the practical implementation across three sections: 
    \begin{itemize}
        \item \emph{instrumentation}, covering the sensors and actuators deployed in the plant. 
        \item \emph{device implementation}, outlining the microcontroller-based panels that interface the instrumentation with the automation network. 
        \item \emph{server-side implementation}, detailing the hardware and software components employed. 
    \end{itemize}

    \subsection{Instrumentation}
    \label{subsec:instrumentation}

        The automation system integrates several industrial-grade inline sensors and actuators relevant for the maple syrup industry. 

        Sensors include \emph{liquid volume}, \emph{flow rate}, \emph{temperature}, \emph{conductivity}, \emph{viscosity}, and \emph{pH}, distributed at strategic points across the process and connected via common industrial interfaces, including 4--20~mA, 2--10~V, digital pulse, RS-232 and Modbus RTU.

        Actuation is primarily achieved through Belimo electric valve actuators, each controlled via 0--10~V position command with corresponding 2--10~V position feedback. 
        A total of 50 valve actuators are deployed, enabling fine-grained and concurrent flow path control across the installation. 
        In addition, three centrifugal pumps are integrated, each driven by relay-based start/stop signals to their contactor panel and instrumented with head pressure sensors and inline flowmeters for monitoring and protection.

    \subsection{Device implementation}

        On the device side, the implementation focuses on providing low-cost, distributed interfaces to sensors and actuators while maintaining uniform communication capabilities over MQTT. 
        Each device is built around a Norvi IIoT ESP32-based industrial controller~\cite{ESP32BasedIndustrial}, integrated in typical industrial DIN-rail enclosures (plastic, to allow the use of integrated antennas) with terminal blocks, allowing fast and flexible building and easy reconfiguration. 
        This controller choice provides a compact and robust form factor and allows the use of the widely used ESP32 platform while exposing all required industrial interfaces. 
        The enclosures operate at 24~VDC and 3~A with both overvoltage and overcurrent protections; under local regulations, these low-voltage and low-current specifications significantly reduce electrical verification costs, since the installation falls under certification exemptions.

\begin{figure}[h!]
			\centering
			\includegraphics[width=\inpaper{0.98\linewidth}{0.95\linewidth}]{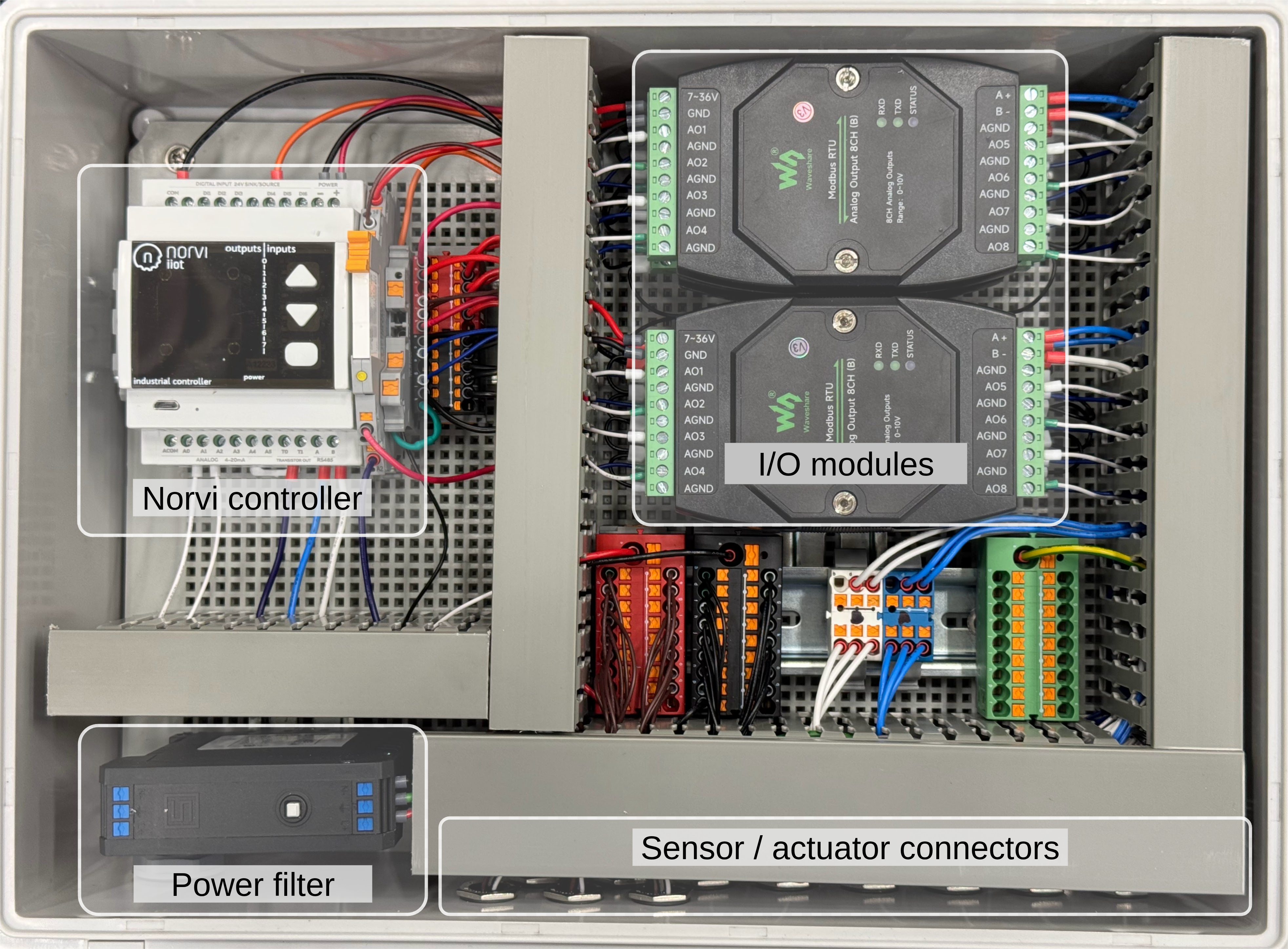}
			\caption{Actuator interface panel example}
			\label{fig:panel}
		\end{figure}

        Sensor panels are built and installed ad-hoc, since their location depends on the specific application. This highlights a strength of the proposed system: modularity at the device level and the use of WiFi networking allow panels to be installed close to the instrumentation -- deployment complexity is low enough to justify installing a panel for every instrument group, which makes for a cleaner "shortest-path" architecture than large monolithic panels. Following this principle and considering the large number of actuators installed in the same location, actuator panels are built as a standard, horizontally-scalable design that meets power limitations mentioned above. Fig.~\ref{fig:panel} shows this design, which supports 16 valve actuators and 1 pump with the following hardware:

        \begin{itemize}
            \item 1 Norvi AE04-I controller (interfacing pump control via relay output, head pressure sensor via 4--20~mA input, flowmeter via digital pulse input)
            \item 16 0--10~V inputs (2x Waveshare Modbus RTU input modules)
            \item 16 0--10~V outputs (2x Waveshare Modbus RTU output modules)
        \end{itemize}

        In total, 8 sensor panels (named "station-2" through "station-9") and 5 actuator panels (named "ctrl-1" through "ctrl-5") are deployed in the plant.

        Firmware is implemented using ESPHome~\cite{openhomefoundationESPHomeSmartHome}, a mature open-source framework with wide peripheral library coverage and an active community, minimizing development effort and ensuring long-term stability. The firmware implements three main features: every datapoint in the system is sampled at 1~Hz and published to MQTT; actuator commands are subscribed from MQTT and executed asynchronously (actuator feedback is included in the aforementioned datapoints); and telemetry data is published once per minute for monitoring purposes. Each device requires only a one-time flashing during commissioning, after which updates can be performed over-the-air (OTA), facilitating maintenance and deployment flexibility.

        While the proposed architecture aims to offload all automation and fault handling logic to the server, attention should be given to how the worst-case scenarios of unrecovered disconnection are be accounted for on the device side. In the present case-study, this has no impact on valves -- the fault lockout system is sufficient and such a disconnection case leaving valves open would have no consequence if the subsystem is isolated. The latter condition, although, requires being able to stop pumps on the device side. This is simply implemented as a "pause", temporarily disabling the pump until connection is restored. If the disconnection is short enough that timeouts haven't been encountered server-side, the operation is resumed; in the opposite case, retained messages from the lockout command will keep the pump disabled on reconnection.

        Each device defined by a unique configuration, but the overall design remains modular and reproducible and templates are shared across devices to maximize reuse. New devices can be commissioned quickly without modification on the server side, and hardware modules can be replaced or scaled as needed without re-engineering the software stack. In combination with the containerized server implementation, this results in a fully modular distributed system where both server- and device-level components can be replicated, extended, or redeployed with minimal effort.

    \subsection{Server-side implementation}

        On the server side, the system implements a microservice architecture with four core open-source components:
        \begin{itemize}
            \item \textbf{Node-RED}~\cite{LowcodeProgrammingEventdriven}: the main component of the architecture.
            \item \textbf{Mosquitto}~\cite{eclipsefoundationEclipseMosquittoOpen}: an industry-standard MQTT broker.
            \item \textbf{QuestDB}~\cite{QuestDBNextgenerationTimeseries}: a high-performance time-series database, used to persist uncompressed device data (sensor and telemetry) as well as internal event data from every layer presented in \S\ref{sec:system-architecture} for later analysis.
            \item \textbf{Grafana}~\cite{grafanalabsGrafanaOpenComposable}: a dashboarding platform for visualization of process data, states, trends, and fault events.
        \end{itemize}

        All services are deployed as \emph{Docker containers} orchestrated with \emph{Docker Compose}. 
        This provides a portable and reproducible environment where each component is versioned and isolated, while inter-service communication is defined declaratively. 
        Only Node-RED and Mosquitto are strictly required for the operation of the proposed system, both of which are stateless outside of their runtime context. QuestDB and Grafana are only implemented as a support for analysis and monitoring, which goes beyond the scope of this paper but is owed the mention that the rich data generated by the system should not be discarded.
        Persistent storage for the two latter services being required, it is managed via Docker volumes.
        The full container configuration is tracked in a dedicated Git branch, allowing reproducible deployments, rollback, and collaborative maintenance.

        This approach provides inherent hardware flexibility: the same configuration can be deployed on industrial servers, virtual machines, or compact embedded platforms without modification. 
        In practice, the current implementation runs on a mid-range x86 server running Debian 12 Linux, as it was available in the plant, but the modular containerized design makes it equally feasible to target constrained environments such as the Raspberry~Pi for lightweight deployments.

        In the full context of the case-study implementation, data integration is handled in parallel with automation. 
        All datapoints published by field devices are ingested directly into QuestDB via Node-RED and the relevant streams are forked in real time into context storage for condition fulfillment checks within operations and routine state-machines. 
        This ensures that control decisions are always based on live process signals while reminding the fact that data persistence needs not be locally handled.
        Beyond automation, this architecture enables real-time processing of operational data for complementary functions such as logistics and traceability monitoring.

\section{Performance evaluation}
\label{sec:evaluation-results}

    The viability of the proposed system is evaluated as part of the case study: (A) quantitavely with an analysis of time constraints and (B) qualitatively with observations of the system's behavior. The system was deployed in a partnered maple syrup boiling center in the Estrie region of Quebec, Canada. Data used for this evaluation was collected over the course of the 2025 production season, which spanned from 2025-03-13 to 2025-04-28. Unless otherwise stated, this entire time period is covered by the results presented.

    \subsection{Time constraints}

        Timing analysis is presented at 3 hierarchical levels: (i) message latency between devices and the server, (ii) command execution metrics from round trips of the actuator control system, and (iii) singular operation run time.

\begin{figure}[h!]
                \centering
                \includegraphics[width=\inpaper{0.99\linewidth}{0.92\linewidth}]{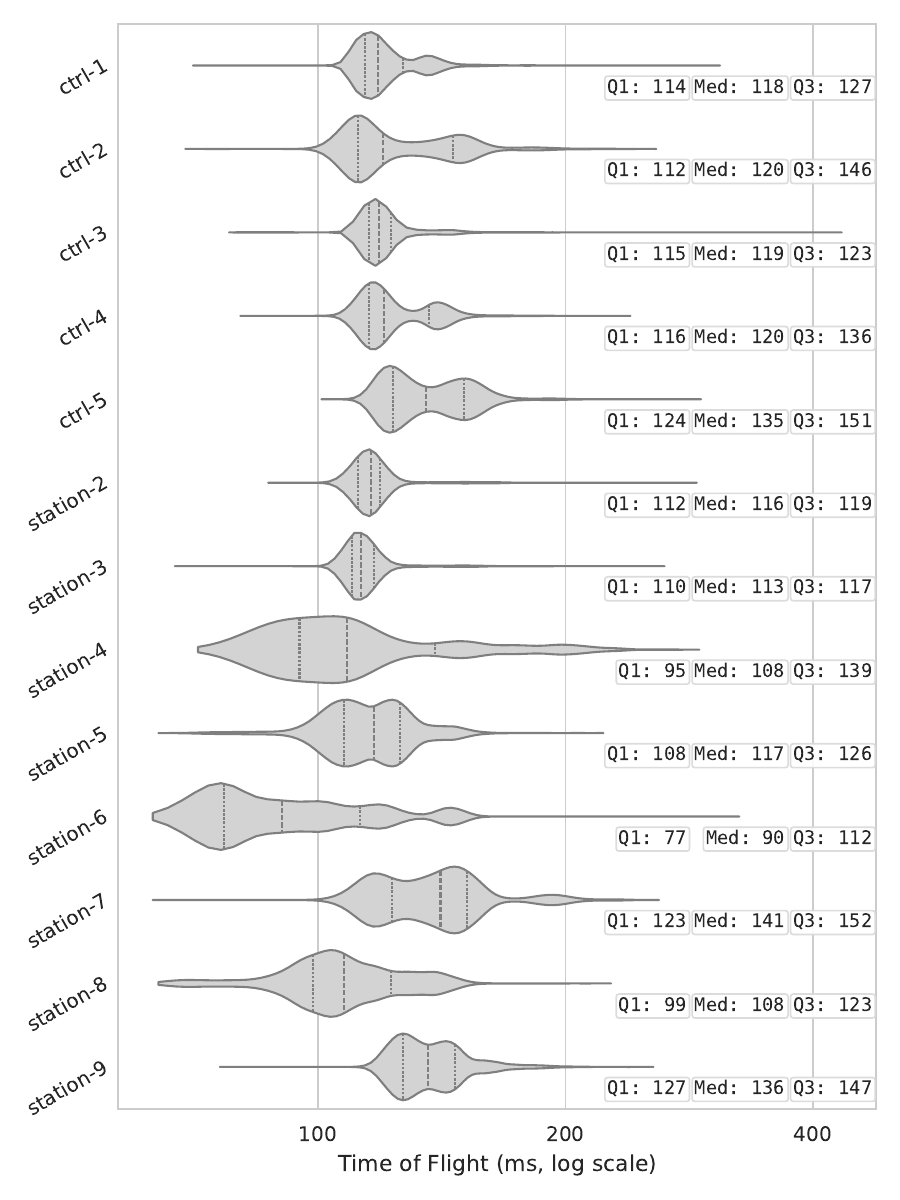}
                \caption{\textbf{Time of flight distribution per device.} Sampled at 1~Hz over 24h ($N=86400$ per device); devices that publish to several topics per sampling interval are averaged such as to have a single measurement per device per sampling interval.}
                \label{fig:tof}
            \end{figure}

        \textbf{(i) Message latency:}
            The time of flight (ToF) of an MQTT message is defined as the time difference between the message being published and the message reception.
            In the present context, the ToF evaluated is the time delta between the timestamp generated by ESPHome (appended to outgoing messages) and a timestamp generated upon reception by the Node-RED server. Both timestamps have a resolution of 1~ms.
            Fig.~\ref{fig:tof} shows the distribution of the ToF for each device.

\begin{figure}[h!]
                \centering
                \includegraphics[width=\inpaper{0.99\linewidth}{0.92\linewidth}]{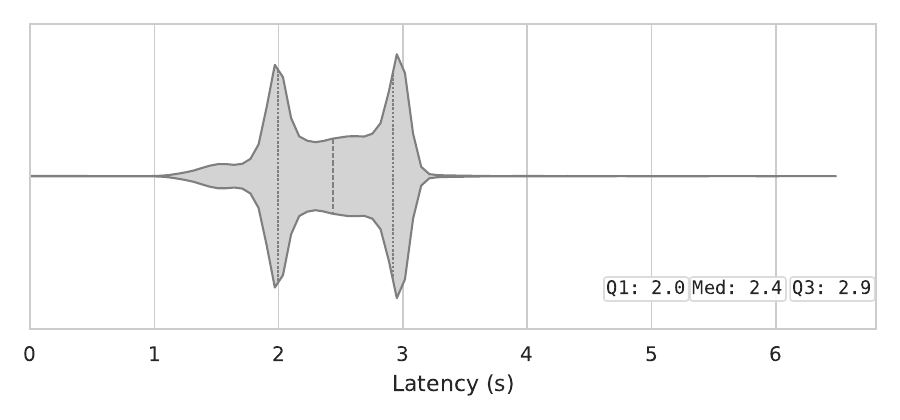}
                \caption{\textbf{Command issuance-to-motion latency distribution.} ($N=28800$) Note the concentrations around the 2 and 3 second marks implicating the 1~Hz feedback rate.}
                \label{fig:latency}
            \end{figure}

        \textbf{(ii) Command execution metrics:}
            The command issuance-to-motion latency (Fig.~\ref{fig:latency}) is the time difference between the command issuance and the reception of the first point of actuator position feedback, both in Node-RED.

\begin{figure}[h!]
                \centering
                \includegraphics[width=\inpaper{\linewidth}{0.92\linewidth}]{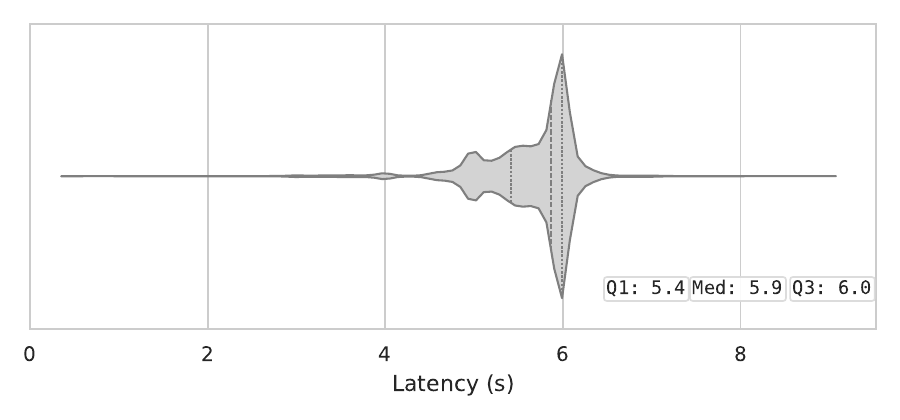}
                \caption{\textbf{Command completion latency distribution.} ($N=28800$) Note the concentration around the 6 second mark implicating the nominal 4-second full-sweep time of the valve actuators and the previously mentioned issuance-to-motion latency.}
                \label{fig:latency_completion}
            \end{figure}

            The command completion time (Fig.~\ref{fig:latency_completion}) is the time difference between the command issuance and the reception of the last point of actuator position feedback (i.e., the point where the actuator has reached the desired position). 

\begin{figure}[h!]
                \centering
                \includegraphics[width=\inpaper{\linewidth}{0.92\linewidth}]{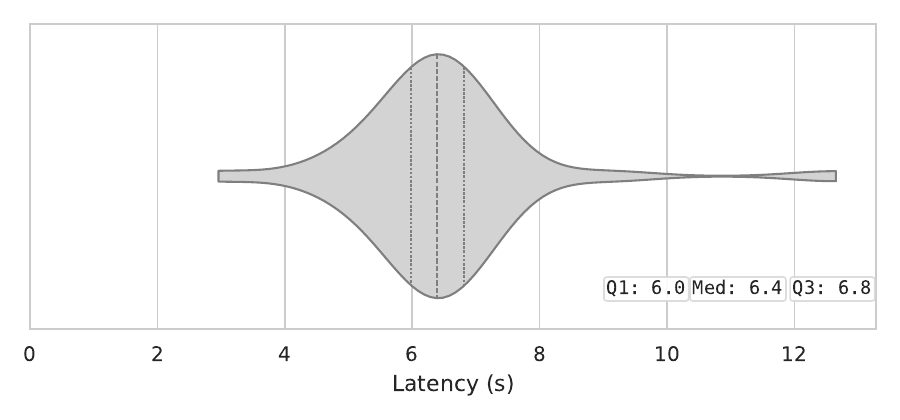}
                \caption{\textbf{Fault lockout time distribution.} ($N=34$) Note the similarity to the command completion latency, and the less concentrated distribution consequent to the smaller sample size.}
                \label{fig:fault_lockout_time}
            \end{figure}

            The fault lockout time (Fig.~\ref{fig:fault_lockout_time}) is the time difference between the fault occurrence and the completion of the command group that isolated the faulted system.

            Overall, these measurements show that the determining factor of the command execution metrics is the actuator run time.

\begin{figure}[h!]
                \centering
                \includegraphics[width=\inpaper{\linewidth}{0.92\linewidth}]{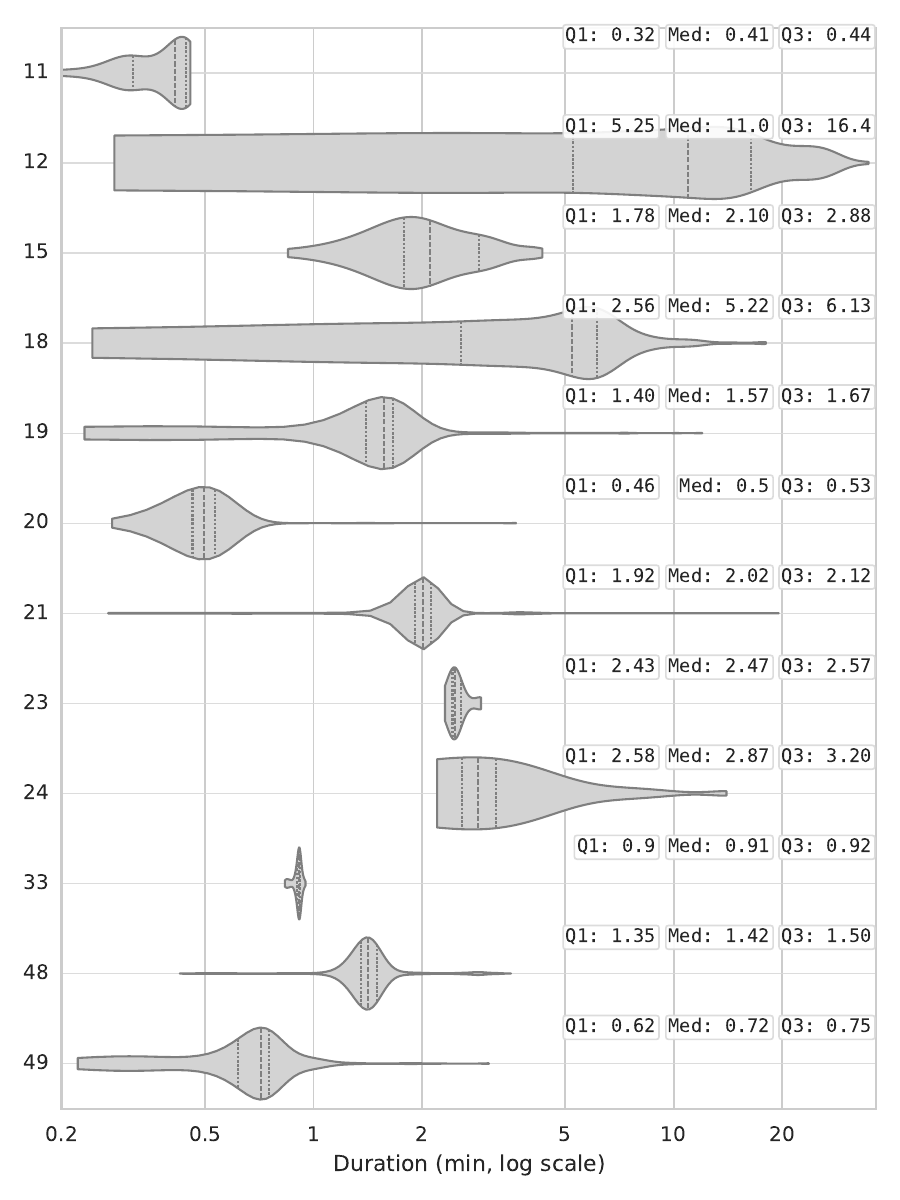}
                \caption{\textbf{Operation run time distribution.} Note the diversity between operations, and the absolute minimum of 12 seconds (sequence of 3 valve actuations with contained condition already fulfilled).}
                \label{fig:operation_run_time}
            \end{figure}

        \textbf{(iii) Operation run time:}
            This metric represents the total run time of singular operations, from interlock grant to release (i.e., not including the time spent in the operation queue, if applicable). Fig.~\ref{fig:operation_run_time} shows the distribution, where the IDs in the y-axis correspond to those shown in Fig.~\ref{fig:op_table}.

        \inpaper{}{\clearpage}
        \textbf{Interpretation:}
            The results show that each layer of the system presents satisfactory performance with timing characteristics at least an order of magnitude smaller than that of the next layer. Specifically, the message ToF is negligible compared to the command execution metrics, implying that the determining factor of the latter is the actuator run time. Moreover, the command execution metrics are also considered negligible compared to the operation run time, implying that both the actuators and the proposed control system are adequately responsive to the specific requirements of the process.

    \subsection{Qualitative observations and discussion}

        During operation, the orchestration executed deterministically and without intervention. Singular operations progressed through lock acquisition, command dispatch, condition fulfillment, and release as designed. The resource-interlock and queueing layer consistently enforced mutual exclusion on shared lines, and queued requests were granted in priority order immediately upon release events. Command retries, escalation paths, and safe-stop fallbacks operated as intended, indicating that actuator abstraction, feedback handling, and completion tracking were sufficient to close the loop in normal operating conditions.

        Fault signaling that did occur was attributable to genuine process anomalies (e.g., empty vessels, flow interruptions external to the control layer) rather than to timing, transport, or orchestration errors. In these instances, the interlock layer isolated the affected resources and propagated structured diagnostics as specified, while unrelated operations continued unaffected. No spurious faults, duplicate alarms, or misrouted commands were recorded. Across the entire season, routine state machines remained monotonic (no unexpected state regressions), idempotent re-invocations produced consistent outcomes, and long-running schedules exhibited stable behavior with no need for manual overrides. These observations, together with the quantitative latency results, support that the proposed system did not limit responsiveness; end-to-end responsiveness was dominated by actuator mechanics, as intended.
        
        As mentioned in \S\ref{sec:implementation-details}, only Mosquitto and Node-RED are strictly required to replicate the system proposed in this paper. On a trial run on a Raspberry Pi 4, these two services together utilized $\approx$25\% of CPU under simulated production load from historical data. In contrast, with the current server used in the case-study, QuestDB alone consumed $\approx$30x more CPU than the Mosquitto/Node-RED pair. Storage size was modest but write-heavy: a mean of $\approx$25 IOPS to disk with $\approx$14 GB growth per operating season. While this falls within the nominal capabilities of SD media, sustained operation at these rates would degrade SD performance and endurance; practical deployments should prefer SSD or offload storage. Consistent with these observations, an edge/cloud split is well suited: keeping Mosquitto and Node-RED at the edge on constrained hardware for low-latency control, and offloading high-IO persistence, exploration, and analytics to a server-class or cloud environment. This model would also facilitate scaling across multiple plants.

\section{Conclusion and future work}
\label{sec:conclusion}

This paper validates an event-driven, IIoT-based architecture as a viable alternative to PLC-centric process control and highlights two specific contributions: 
\begin{enumerate}
    \item a modular device abstraction layer that decouples control logic from hardware implementation and supports modular horizontal scaling and easy reconfiguration;
    \item an asynchronous interlock with priority queueing that allows safe, reusable runtime arbitration of shared resources, removing the need for systematic enumeration of conflict cases at design time.
\end{enumerate}
Together, they reduce design workload and integration cost while preserving determinism, safety, and fault isolation.

Results show that the system exhibits adequate performance with regards to timing characteristics, where the main defining factors are process-specific rather than limitations of the proposed system. Qualitative observations show that the system operated deterministically and without intervention, and that the system was able to handle faults as intended.

As to not constrain the viability of the proposed system to processes with "loose" timing requirements such as the one presented in the case-study, future work could investigate the application of the proposed system in a process with more stringent requirements. This could be done with a \textit{hybrid} configuration implementing some software-defined automation logic on-device to comply with the timing constraints, while keeping the rest of the logic in Node-RED and wrapping the new device-level logic as an actuator abstraction.

Future work could also distribute the architectural primitives (resource interlock and priority queue, actuator abstraction layer, and routine templates) as open-source Node-RED nodes and subflows with examples to facilitate reuse. Another contribution could be to implement and benchmark an edge/cloud split in which automation executes on the edge (e.g., Raspberry Pi) and data persistence, visualization, and analytics are offloaded to cloud services.

\bibliographystyle{IEEEtran}
\bibliography{refs}

\begin{IEEEbiography}[{\includegraphics[width=1in,height=1.25in,clip,keepaspectratio]{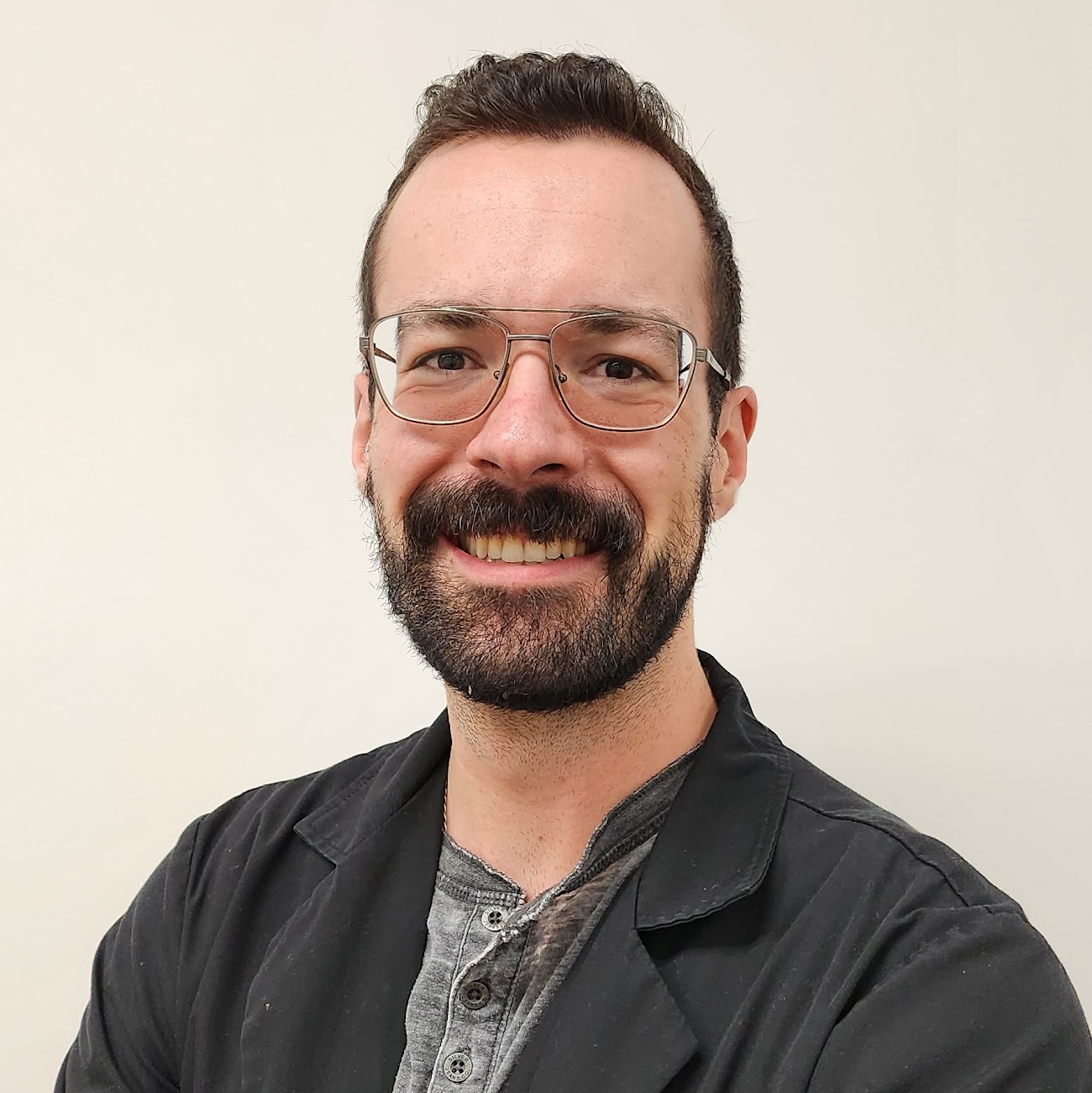}}]{Thomas Bernard} received the B.Eng. degree in computer engineering in 2018, the M.Sc. degree in electrical engineering in 2021, and is currently pursuing the Ph.D. degree in electrical engineering, at Université de Sherbrooke, Sherbrooke, QC, Canada. His major field of study is Industry~4.0 automation systems.

He has worked as an R\&D Consultant in industrial IoT since 2015, with roles spanning embedded systems, distributed automation, and data integration. He is currently R\&D Coordinator at AFCA, La Patrie, QC, and a Lecturer in Robotics Engineering at Université de Sherbrooke. He is currently a Candidate to the Engineering Profession (CEP) with the Ordre des Ingénieurs du Québec.
\end{IEEEbiography}

\begin{IEEEbiography}[{\includegraphics[width=1in,height=1.25in,clip,keepaspectratio]{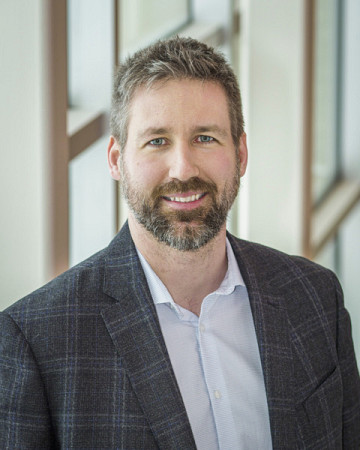}}]{François Grondin} (Member, IEEE) received the
    B.Sc. degree in electrical engineering from McGill
    University, Montreal, QC, Canada, in 2009, and the
    M.Sc. and Ph.D. degrees in electrical engineering
    from the Université de Sherbrooke, Sherbrooke,
    QC, Canada, in 2011 and 2017, resepectively.
    After completing postdoctoral work with the
    Computer Science and Artificial Intelligence
    Laboratory, Massachusetts Institute of Technology,
    Cambridge, MA, USA, in 2019. He became a
    Faculty Member with the Department of Electrical
    Engineering and Computer Engineering, Université de Sherbrooke. He is currently a Member with the Ordre des ingénieurs du Québec.  His research interests include robot
    audition, sound source localization, speech enhancement, sound classification,
    and machine learning.
\end{IEEEbiography}

\begin{IEEEbiography}[{\includegraphics[width=1in,height=1.25in,clip,keepaspectratio]{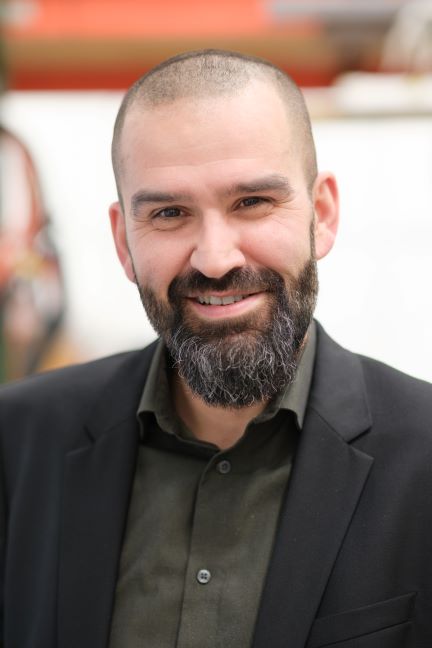}}]{Jean-Michel Lavoie}
    received the B.Sc. degree in chemistry and the M.Sc. and Ph.D. degrees in wood sciences from Laval University, Québec, Canada. Prof.~Lavoie is the founder and director of the Biomass Technology Laboratory, leading a team of 30--50 researchers dedicated to renewable energy and bioprocessing. His group collaborates with numerous industrial partners in Québec and abroad.
\end{IEEEbiography}

\end{document}